\documentclass{iau} 
\usepackage{graphicx}

\title[Dusty Galaxies at the Highest Redshifts] 
{Dusty Galaxies at the Highest Redshifts}

\author[David L Clements et al.]   
{David L Clements$^1$, Josh Greenslade$^1$, Dominik A. Riechers$^2$, Julie Wardlow$^{3,4}$, Ismael P{\'e}rez-Fournon$^{5,6}$, The HerMES Red Collective, The HerMES \& H-ATLAS Consortia }

\affiliation{$^1$Physics Department, Blackett Lab, Prince Consort Road, London SW7 2AZ, UK\\[\affilskip]
$^2$Department of Astronomy, Cornell University, Space Sciences Building, Ithaca, NY 14853, USA\\[\affilskip]
$^3$Dark Cosmology Centre, Niels Bohr Institute, University of Copenhagen, Denmark\\[\affilskip]
$^4$Centre for Extragalactic Astronomy, Department of Physics, Durham University, South Road, Durham, DH1 3LE, UK\\[\affilskip]
	$^{5}$Instituto de Astrof\'{i}sica de Canarias, C/ V\'{i}a L\'{a}ctea, E-38200 La Laguna, Tenerife, Spain\\
	[\affilskip]
    	$^{6}$Departimento de Astrof\'{i}sica, Universidad de La Laguna, E-38206, La Laguna, Tenerife, Spain\\
[\affilskip]
}

\pubyear{2015}
\volume{319}  
\jname{Galaxies at High Redshift and Their Evolution over Cosmic Time}
\editors{S. Kaviraj, ed.}
\begin{document}

\maketitle

\begin{abstract}
We show that the use of red colour as the basis for selecting candidate high redshift dusty galaxies from surveys made with {\em Herschel} has proved highly successful. The highest redshift such object, HFLS3, lies at $z=6.34$ and numerous other sources have been found. Spectroscopic followup confirms that most of these lie at $z>4$. These sources are found in such numbers that they represent a challenge to current models of galaxy evolution. We also examine the prospects for finding dusty galaxies at still higher redshifts. These would not appear in the SPIRE surveys from {\em Herschel} but would be detected in longer wavelength, submm, surveys. Several such `SPIRE-dropouts' have been found and are now subject to followup observations.

\end{abstract}

\firstsection 
\section{Introduction}

Deep surveys at optical and near-IR wavelengths have revealed much about the formation and evolution of galaxies - see for example the review papers by Finkelstein, Devriendt and Fall in the current volume. The discovery of the Cosmic Infrared Background (CIB; Puget et al., 1996; Fixsen et al., 1998), however, demonstrated that rest-frame optical and UV emission did not tell us the whole story. Instead, it showed that roughly half the optical/UV light generated in the history of the universe was absorbed by dust which heated, and re-radiated this energy in the rest-frame far-IR ie. at wavelengths peaking at about 100$\mu$m. The SCUBA observations of the Hubble Deep Field (Hughes et al., 1998) showed that the far-IR luminous galaxies responsible for the CIB behave in a very different way to the optical sources in this field. Whereas HST detected about 2000 galaxies, SCUBA found just five. Moreover, it took 15 years to properly identify and measure the redshift of the brightest of these, HDF850.1 (Walter et al., 2012) since it was so faint in the optical/near-IR that it was undetected in the HDF images. With a redshift of 5.18, measured using far-IR emission lines, this source, and the many others that have come to light since HDF850.1 was first detected (e.g. Scott et al., 2011; Harris et al., 2012; Lupu et al., 2012),  demonstrate the need for far-IR and submm surveys to complement optical surveys so that we can have a complete picture of galaxy formation and evolution. These surveys became possible with the launch of the {\em Herschel Space Observatory}.

\begin{figure}
\begin{center}
 \includegraphics[width=4.5in]{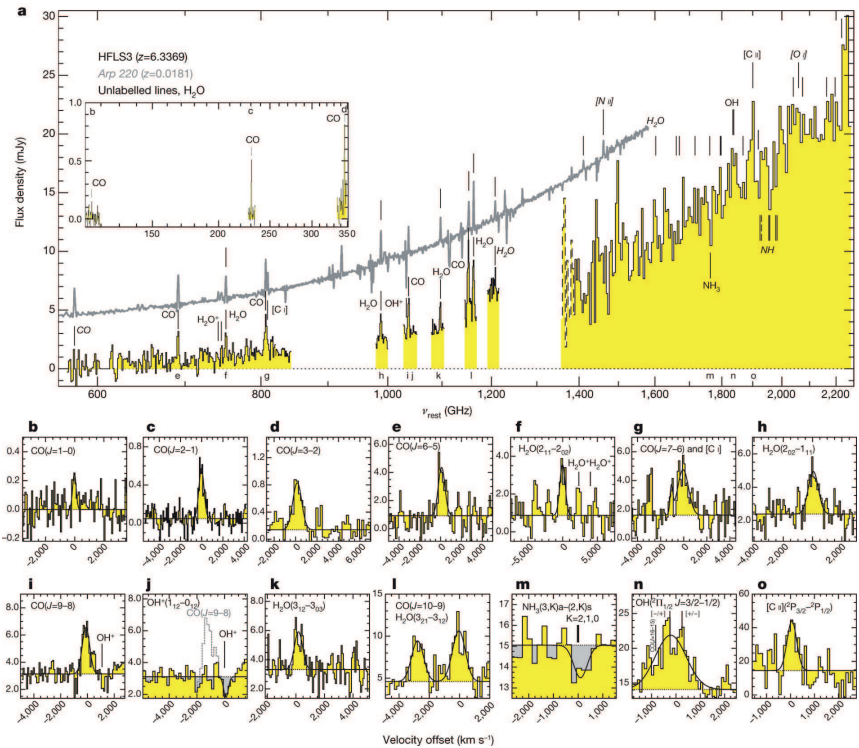} 
 \caption{The spectrum of HFLS3, the current highest redshift dusty starburst galaxy known, at $z=6.32$. Shown here are both wide band spectroscopy (a) from a combination of observations using CARMA, JVLA, PdB and Z-spec and (b to o) detailed spectra of specific lines including species of CO, H$_2$O, OH, OH$^+$, NH$_3$, [CI], and [CII]. Figure from Riechers et al. (2013), from which more details can be obtained.}
   \label{fig1}
\end{center}
\end{figure}

\section{High Redshift Searches with Herschel}

{\em Herschel} (Pilbratt et al., 2010) conducted several large area surveys of the extragalactic sky using the PACS (Poglitsch et al., 2010) and SPIRE instruments (Griffin et al., 2010). The largest of these are the H-ATLAS survey (Eales et al., 2010) and the HerMES survey and its related extensions (Oliver et al., 2012). Together these projects provide a total sky coverage of about 1000 sq. deg., to varying sensitivities, at wavelengths of 100, 160 (PACS; Albrecht et al., 2010), 250, 350 and 500$\mu$m (SPIRE; Griffin et al., 2010). Since the dust Spectral Energy Distribution (SED) of a galaxy typically peaks around 100$\mu$m, we can use the colours of sources in these bands to select candidate high redshift objects. A source whose SED is rising from 250 to 350 to 500$\mu$m, for example, is likely to lie at $z>4$ unless its dust temperature is substantially lower than is seen in local equivalents. Contamination of this selection by non-thermal sources is small, and easily eliminated by consulting radio catalogues. Searches for such `500-riser' sources have so far been conducted over a total of about 300 square degrees in the HerMES survey (Dowell et al., 2014; Asboth, 2015), yielding about 500 candidates. Followup observations have so far secured spectroscopic redshifts for 13 of these sources, 10 of which are at $z>4$ (and all at $z>3$; Riechers et al., in prep). Examples of these sources include the $z=6.34$ source HFLS3 (Riechers et al., 2013), the highest redshift source found so far (the spectrum of which is shown in Fig. \ref{fig1}), and the strongly lensed $z=5.29$ source HLock-102. While some sources, like HLock-102, are strongly lensed, even when lensing amplification is taken into account all these objects are forming stars at very high rates, $>1000M{_\odot}$/yr. These objects likely represent the formation of what will become a giant elliptical galaxy (see eg. Farrah et al., 2006). The overall number density of these sources is significantly above that predicted by the latest generation of galaxy evolution models (Dowell et al., 2014; Asboth et al., 2015), so understanding this population is clearly important for improving our knowledge of the formation and evolution of the most extreme galaxies.

Increasing the number of spectroscopically confirmed high redshift dusty galaxies is a necessary step towards this goal. Since these systems are very faint in the optical/near-IR, spectroscopic confirmation has so far been done in the submm, typically looking for redshifted CO or [CII] lines. Most submm receivers currently available, including those on ALMA, have relatively narrow bandwidths, so blind line searches for redshift determination require multiple observations at different tunings. This can lead to significant observing overheads. A new generation of submm instruments, however, is gradually becoming available which offer much larger instantaneous bandwidths. These include SWARM on the SMA, which will provide 32GHz bandwidth (Weintroub et al., 2014), the 38GHz RSR on the LMT (Chung et al., 2009) and the 32GHz EMIR on the IRAM 30m (Carter et al., 2012), and the 32GHz bandwidth planned for NOEMA, as well as more specialised instruments such as Z-Spec (Lupu et al., 2012) and Zpectrometer (Harris et al., 2012).

\section{Higher Redshift Searches}

Our success in finding unexpectedly large numbers of $z>4$ dusty sources using the simple `500-riser' selection technique begs the question of whether there are still higher redshift dusty galaxies that we can find. The discovery of a massive, dusty galaxy forming stars at a moderate rate at $z\sim7.5$ (Watson et al., 2015) argues that such objects should exist. However,
the steep drop off in dust emission at rest-frame wavelengths shortward of the peak of the SED, which allows us to identify `500-risers' as candidate $z>4$ sources, means that galaxies at redshifts significantly higher than $z\sim$6 will be too faint to be seen by SPIRE. They will, nevertheless, be detectable by longer wavelength submm observations and would thus appear as submm sources that lack counterparts in SPIRE observations. Comparison of sources found in deep submm observations with matching {\em Herschel} data can then identify such `SPIRE dropouts' for further confirmation and study with followup observations.

\begin{figure}
\begin{center}
 \includegraphics[width=5in]{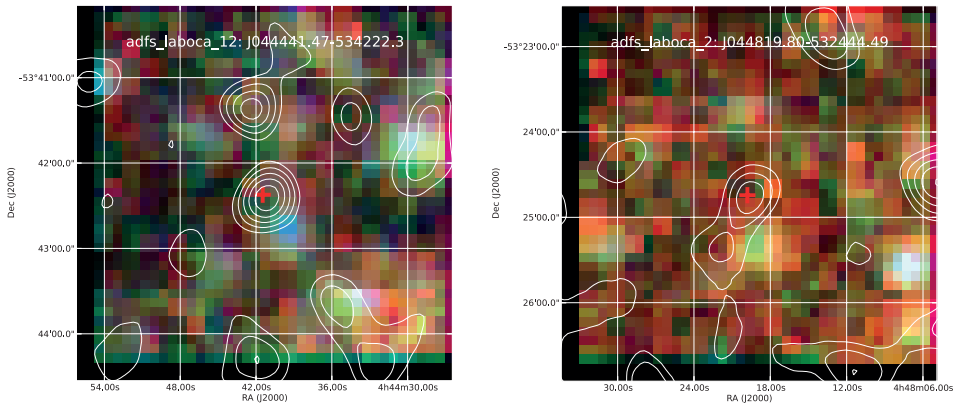} 
 \caption{Two SPIRE dropouts found in the ADF-S region. Contours are AzTEC 1.1mm (Hatsukae et al., 2011) and the image is a 3 colour rendition (250, 350 and 500$\mu$m as blue, green, and red respectively) of the SPIRE data. Note the absence of any SPIRE detections associated with these sources, both of which are seen by AzTEC at $>5\sigma$.}
   \label{fig2}
\end{center}
\end{figure}

We have undertaken searches for SPIRE dropouts using SCUBA2 data collected in parts of the H-ATLAS survey region and with AzTEC data taken in the ADF-S field (Hatsukade et al., 2011) which was observed by HerMES. A number of candidate dropouts were found, two of which are shown in Fig. \ref{fig2}. These sources are now subject to a variety of followup observations to better constrain their submm SEDs and to measure their redshifts spectroscopically. Observations of one of these sources with the SMA and with the NIKA instrument on the IRAM 30m confirm that this source is likely at $z>5$. Followup observations of these sources, including spectroscopy to unambiguously measure their redshifts, and the next step in this programme.

\section{Conclusions}

The use of far-IR/submm colours to identify candidate $z>4$ dusty galaxies in {\em Herschel} surveys has led to the discovery of a population of rapidly starforming dusty high redshift galaxies. These sources exist in larger numbers than the current generation of galaxy evolution model can predict, so studies of these objects can provide new insights into galaxy formation and evolution. These colour selection methods can be extended to allow the search for higher redshift sources by seeking `SPIRE dropouts' in surveys at longer, submm wavelengths. Our first searches for such objects have found a number of candidate high redshift sources which are the subject of followup observations.

\end{document}